\documentclass[preprint,amsmath,amssymb,aps,prb]{revtex4-1}
\usepackage{epsfig}
\usepackage{graphicx}
\usepackage{dcolumn}
\usepackage{bm}
\usepackage{indentfirst}
\usepackage{latexsym}
\usepackage{multirow}
\let\Gamma\varGamma

\begin{document}

\title{Transition-metal dichalcogenide bilayers: switching materials for spin- and valleytronic applications}
\author{Nourdine Zibouche,$^{1,2}$ Pier Philipsen,$^2$ Agnieszka Kuc,$^1$ and Thomas Heine$^1$}
\email{t.heine@jacobs-university.de}
\affiliation{$^1$School of Engineering and Science, Jacobs University Bremen, Campus Ring 1, 28759 Bremen, Germany\\
$^2$Scientific Computing \& Modelling NV, De Boelelaan 1083, 1081 HV Amsterdam, The Netherlands}

\date{}
\begin{abstract}
We report that an external electric field applied normal to bilayers of transition-metal dichalcogenides TX$_2$, M = Mo, W, X = S, Se, creates significant spin-orbit
splittings and reduces the electronic band gap linearly with the field strength.
Contrary to the TX$_2$ monolayers, spin-orbit splittings and valley polarization are absent in bilayers due to the presence of inversion symmetry. This symmetry can
be broken by an electric field, and the spin-orbit splittings in the valence band quickly reach similar values as in the monolayers (145 meV for MoS$_2$ $\ldots$
418 meV for WSe$_2$) at saturation fields less than 500 mV \AA$^{-1}$. The band gap closure results in a semiconductor--metal transition at field strength between
1.25 (WX$_2$) and 1.50 (MoX$_2$) V \AA$^{-1}$. Thus, by using a gate voltage, the spin polarization can be switched on and off in TX$_2$ bilayers, thus activating
them for spintronic and valleytronic applications. 
\end{abstract}

\maketitle

\section{Introduction}
Two-dimensional (2D) materials have been intensively investigated in the past few years for their applications in next-generation nanoelectronics, including
spintronics\cite{Zutic2004} and valleytronics.\cite{Cao2012} Transition-metal chalcogenides (TMCs) of the form TX$_2$ (T=Mo, W, X=Se, S) are of particular interest,
as they have several interesting intrinsic properties, such as direct band gaps\cite{Mak2010, Splendiani2010, Kuc2011} or giant spin-orbit (SO) coupling 
(SOC)\cite{Zhu2011, Sun2013} in monolayered (ML) forms. Excellent electronic properties of TMCs have recently led to the production of first nanoelectronic devices
based on TMC-MLs, including thin film transistors, logical circuits, amplifiers and photodetectors.\cite{Wang2012, Lopez-Sanchez2013, Radisavljevic2011, Radisavljevic2012}
It has also been reported that external stimuli, e.g.\ tensile strain,\cite{Yun2012, Scalise2012, Ghorbani2013, Ghorbani2013a} can strongly influence the electronic properties of TX$_2$ layers.

SOC is a relativistic effect that occurs for honeycomb 2D lattices with a broken inversion symmetry, such as in 2$H$ TMC-MLs. Thus, for MoS$_2$, a giant spin-orbit-induced
band splitting of $\sim$100 meV was reported from Raman experiments.\cite{Sun2013} In agreement, from first-principles the SO splitting in TX$_2$ MLs was calculated to 
be in the range 148--480 meV, with the limiting values for MoS$_2$ and WTe$_2$ MLs, respectively.\cite{Zhu2011,Zibouche2014} At the same time, TX$_2$ monolayers are very 
stable with respect to external electric fields, with a semiconductor-metal transition being reported for fields stronger than 4 V \AA$^{-1}$.\cite{Zibouche2014} In contrast,
TX$_2$ bilayers have been found to be much more sensitive to external electric fields, with band gaps reducing linearly with respect to the external field.\cite{Ramasubramaniam2011,QLiu2012}

It is very interesting to note that the SO splitting in TX$_2$ monolayers disappears nearly completely when going to bilayers (BLs). 
Indeed, in spin- and valleytronic applications, it may be useful to have a material where the polarization can be switched on and off. This can be achieved if the 
inversion symmetry in the bilayer is broken by an external factor, most conveniently by an electric field normal to the lattice plane.
It has already been suggested that the inversion symmetry can be broken in MoS$_2$ BLs through an external electric field applied normal to the planes, which 
leads to a potential difference between individual layers, and allows the control of valley polarization.\cite{Wu2013} This effect should be even more pronounced
for TX$_2$ materials that show stronger spin-orbit splittings in the monolayers, i.e. for WS$_2$ and
WSe$_2$.\cite{Xu2014} Yuan et al.\cite{Yuan2013} have investigated the out-of-plane Zeeman-type spin polarization in WSe$_2$ bilayer-based transistor using ionic-liquid-gate voltage.
They have shown that such spin splitting can be induced and modulated by a perpendicular external electric field. 

Therefore, we investigate here in detail the electronic structure of TX$_2$ bilayers as function of an external electric field. We will show that the same SO splittings
can be achieved in TX$_2$ bilayers as in the corresponding monolayers, for the field strength in the range of 200-600 mV \AA$^{-1}$. At this field strength, the materials
are still semiconductors with appreciable band gaps of more than 500 meV. However, the band gap is a linear function of the applied electric field and this provides 
additional means to tune the electronic properties.
We have found that the field strength of about 1.5 V \AA$^{-1}$ is sufficient for the semiconductor -- metal transition.
This electric field strength can be achieved experimentally using, e.g.\ ionic liquid gating.\cite{Ye2012, Ye2013, Zhang_Ye2013}

\section{Methods}

All calculations were carried out using density-functional theory (DFT) with the PBE\cite{PBE} exchange-correlation functional, with added London dispersion corrections 
as proposed by Grim\-me,\cite{Grimme2006} and with Becke and Johnson damping (BJ-damping) as implemented in the ADF/BAND package.\cite{BAND1,BAND} 
Local basis functions (numerical and Slater-type basis functions of valence triple zeta quality with one polarization function (TZP)) were adopted for all atom types,
and the frozen core approach (small core) was chosen. The $k$-point mesh over the Brillouin zone was sampled according to the Wiesenekker-Baerends scheme,\cite{ADF1}
where the integration parameter was set to 5, resulting in 15 $k$-points in the irreducible wedge. All TMC-BL structures (atomic positions and lattice vectors) are 
fully optimized including scalar relativistic (SR) corrections, which are expressed by the Zero Order Regular Approximation (ZORA)\cite{Pier1997,E_v_Lenthe1993,E_v_Lenthe1999,Filatov2003} to 
the Dirac equation. The implementation of the analytical gradients for SR-ZORA is based on a modiﬁcation of the energy gradients implementation in the non-relativistic case.
The difference to the latter arises in the calculation of the kinetic energy gradients. Moreover, the full relativistic ZORA includes both the SR-ZORA and spin-orbit
interactions.\cite{E_v_Lente1996,E_v_Lenthe1999} The maximum gradients threshold was set to $10^{-4}$ Hartree \AA$^{-1}$.
The lattice parameters and interlayer spacings are given in Table 1.
To obtain electronic structures (band structures and resulting electronic band gaps and spin-orbit splittings) at these optimized coordinates we performed
full relativistic ZORA calculations.\cite{Pier1997,E_v_Lente1996}
At the same level of theory, the response to an external electric field normal to the lattice planes, ranging from 0.0--1.5 \AA$^{-1}$, has been calculated.
In the ADF/BAND, the static electric field is homogeneous and implemented along the $z$-direction (i.e.\ the non-periodic direction).
It is important to note that neither the applied electric fields nor SOC do influence the BL geometries. 
\begin{table}[ht!]
\renewcommand{\thetable}{\arabic{table}}
\centering
\caption{\label{tab:1} Calculated lattice parameters $a$ (in \AA) and the interlayer distances, $d$, measured between the metal planes (in \AA) of all TMC bilayers at the equilibrium. 
The corresponding values from Ref.\cite{Ramasubramaniam2011} are given for comparison. Note that the authors used fixed experimental structures. The reoptimized values for MoS$_2$ BL 
from Ref.\cite{Ramasubramaniam2011} is given in parethesis.}
\medskip
\begin{tabular}{c|cc|cc}
\hline
\textbf{System} & \textbf{$a$} & \textbf{$d$} & \textbf{$a$} & \textbf{$d$} \\
\hline
MoS$_2$ & 3.155 & 6.122 &  3.160 (3.199) & 6.147 (6.180) \\
MoSe$_2$ & 3.268 & 6.425 &  3.299 & 6.469 \\
WS$_2$ & 3.147 & 6.147 & 3.153 & 6.162 \\
WSe$_2$ & 3.266 & 6.413 &  -- & --\\
\hline
\end{tabular}
\end{table}

\section{Results and Discussion}

\begin{figure}[h!]
\begin{center}
\includegraphics[width=0.75\textwidth,clip]{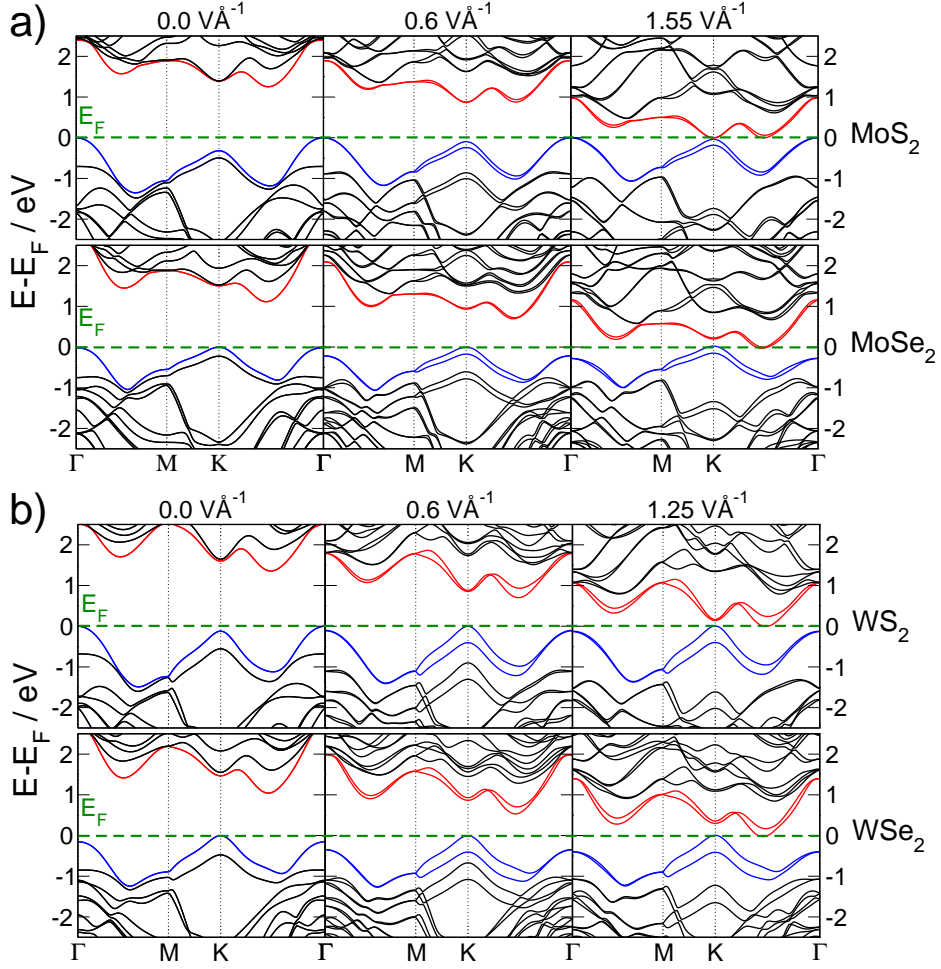}
\caption{\label{fig:bs} Calculated band structures of TX$_2$ bilayers under external electric field. Fermi level (E$_{\rm{F}}$) is shifted to the top of valence band.}
\end{center}
\end{figure}

Calculated band structures of all studied TX$_2$ materials in the presence of external fields of 0.00, 0.60 and 1.55 V \AA$^{-1}$ are given in Fig.~\ref{fig:bs}. 
In the absence of the electric field, the results are in close agreement with values reported earlier for the TX$_2$ bilayers.\cite{Kuc2011}
All systems are indirect band gap semiconductors with band gaps ($\Delta$) of 1.26, 1.14, 1.36, and 1.07 eV for Mo$_2$, MoSe$_2$, WS$_2$, and WSe$_2$, respectively.
For the sulphide BLs, Kuc et al.\cite{Kuc2011} have obtained $\Delta$ of about 1.48 from the PBE non-relativistic calculations, while Dashora et al.\cite{Dashora2013} calculated $\Delta$ of MoS$_2$ to be about 1.27 eV using the FP-LAPW approach with Wu and Cohen exchange-correlation potential.
Using local density approximation within plane-waves approach, Terrones et al.\cite{Terrones2013} obtained band gaps of BLs of 1.11, 1.05, 1.36, and 1.29 eV for Mo$_2$, MoSe$_2$, WS$_2$, and WSe$_2$, respectively.
The valence band maximum (VBM) is located at the high-symmetry $K$ point of the Brillouin zone (BZ) for the selenides,
while it is found at the $\Gamma$ point for the sulphides. The conduction band minimum (CBM) is always located at a low-symmetry point between $K$ and $\Gamma$ (Fig.~\ref{fig:bs}, left panel).

The external electric field polarizes the electron density and thus introduces an anisotropy which creates an appreciable spin-orbit (SO) coupling (SOC).
SO splitting is observed in both conduction and valence bands, with the latter ones being more pronounced (Fig.~\ref{fig:bs}, middle panel).
The SO splitting of the VBM appears to have a natural saturation with a value very close to that of the respective monolayer.
This saturation is reached already at rather small inversion symmetry breaking caused by field strength as small as 200 mV \AA$^{-1}$ (see Table 2).
At the same time, the valence bands are shifted closer to the Fermi level and thus the band gap is reduced. 
For larger fields of 1.55 V \AA$^{-1}$ for MoX$_2$ and 1.25 V \AA$^{-1}$ for WX$_2$, the conduction and valence bands cross the Fermi level and the systems become metallic (Fig.~\ref{fig:bs}, right panel).

\begin{figure}[h!]
\begin{center}	
\includegraphics[width=0.45\textwidth,clip]{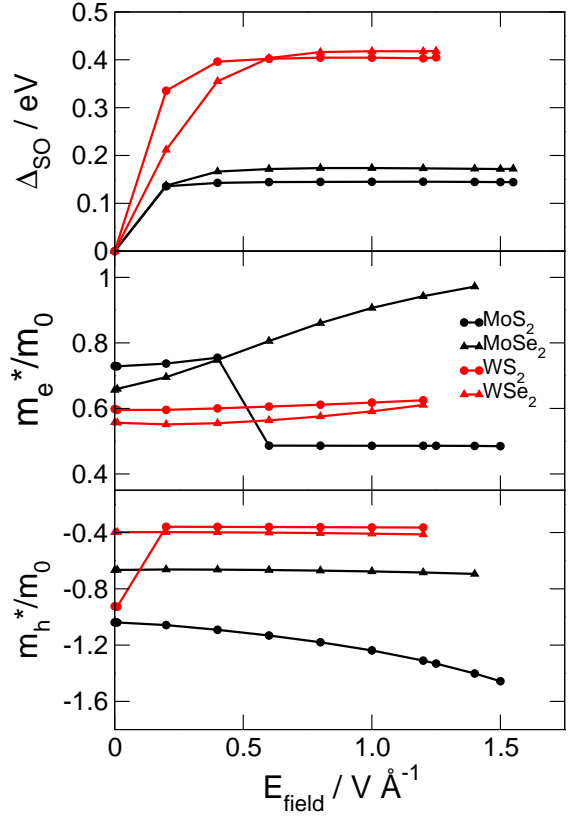}
\caption{\label{fig:em} Calculated valence band spin-orbit splitting values ($\Delta_{SO}$) and the effective masses of electrons and holes ($m_e^*$ and $m_h^*$) of the TX$_2$ bilayers
as function of applied perpendicular electric field. $\Delta_{SO}$ is calculated at the $K$ point, while the effective masses are obtained from the valence band maxima and conduction 
band minima for the holes and electrons, respectively. Note the negative scale on the $y$-axis for hole effective masses that arises from the curvature of the valence band maximum.}
\end{center}
\end{figure}

Our results show an appreciable Stark effect\cite{Stark1915}, that is, due to SOC in an external electric field spin splittings are induced to the electronic bands in the structurally centro-symmetric BLs.
The external field polarizes the electrons in the BLs in such a way that the inversion symmetry is broken. 
This causes SO splittings in a similar way as in the monolayers. Fig.~\ref{fig:em} shows the spin splitting values $\Delta_{SO}$ at the $K$ point for the whole range of applied field strengths. 
At zero field strength, the spin-orbit splitting in BLs is zero, however, for very weak fields the situation drastically changes.
In the valence band maximum (VBM), $\Delta_{SO}$ reaches its maximum of 170 meV (420 meV) for molybdenum (tungsten) dichalcogenides and stays unchanged for the whole range of applied field strengths (see Table 2).
This is in close agreement with $\Delta_{SO}$ reported by Ramasubramaniam et al.\cite{Ramasubramaniam2011} for MoS$_2$ BLs, who reported $\Delta_{SO}$ = 140 meV at similar field strength.
In the CBM of all TX$_2$ bilayers we have obtained non-zero $\Delta_{SO}$, however, the values are smaller than those in the VBM.
The much larger $\Delta_{SO}$ accounted for the WX$_2$ BLs are due to the heavier tungsten atoms. Small variations in $\Delta_{SO}$ due to atom mass differences 
can be observed between selenides and sulphides, but they do not exceed 30 meV.
Incidently, the calculated spin-orbit splittings of the valence bands almost coincide with the values known for the corresponding TX$_2$ monolayers.\cite{Zhu2011, Sun2013, Zibouche2014}

\begin{table}[ht!]
\renewcommand{\thetable}{\arabic{table}}
\centering
\caption{\label{tab:2} The calculated band gap values at zero electric field ($\Delta_{E=0}$), the critical electric fields for the semiconductor-metal transition (E$_{crit}$), 
the spin-orbit splitting values saturated ($\Delta_{SO}^{sat}$) and at the electric field of 200 mV \AA$^{-1}$ ($\Delta_{SO}^{E=0.2}$), and the electric field values for the SO saturation of 90\% (E$_{SO=90\%}$).}
\medskip
\begin{tabular}{c|ccc|cc}
\hline
\multirow{2}{*}{\textbf{System}} & \textbf{$\Delta_{E=0}$ } & \textbf{E$_{crit}$} & \textbf{$\Delta_{SO}^{sat}$} & \textbf{$\Delta_{SO}^{E=0.2}$} & \textbf{E$_{SO=90\%}$} \\
& (eV) & (V \AA$^{-1}$) & (meV) & (meV) &  (V \AA$^{-1}$)\\
\hline
MoS$_2$       &   1.26 & 1.50 & 145 & 136 & 0.2\\
MoSe$_2$     &   1.14 & 1.50 & 173 & 137 & 0.4\\
WS$_2$	      &  1.36 & 1.25 & 404 & 335 & 0.4\\
WSe$_2$       &   1.07 & 1.25 & 418 & 212 & 0.6\\
\hline
\end{tabular}
\end{table}

Interestingly, the effect of the external electric field on the mobilities of electrons and holes is quite different in the four TX$_2$ structures (Fig.~\ref{fig:em}). For MoS$_2$, effective hole masses 
$m_h^*$ increase with the applied field due to flattening of the bands. As the CBM moves already at small applied field (400 mV \AA$^{-1}$) to the $K$ point in the BZ, we observe a discontinuity of the 
effective electron masses $m_e^*$ and assume a strong increase in electron mobility for this system.
In MoSe$_2$, the $m_e^*$ increase with applied electric field, while the $m_h^*$ remain stable. Both tungsten dichalcogenides show stable $m_e^*$ with respect to the external field. The same is true for 
the $m_h^*$ in WSe$_2$, while for WS$_2$ VBM changes from $\Gamma$ to $K$ in the BZ at fields higher than 200 mV \AA$^{-1}$. In general, the effective masses are smaller for the tungsten compounds, with
the exception of $m_e^*$ for MoS$_2$ at high external fields.  

Two reports have already discussed an interesting evolution of the electronic band gap $\Delta$ as function of the external field.
Ramasubramaniam and co-workers\cite{Ramasubramaniam2011} have shown that the electronic structure of MoX$_2$ (X=S, Se, Te) and WS$_2$ bilayers can be influenced with a perpendicular electric field.
Their first-principles based plane wave simulations suggested that the electronic band gaps decrease linearly with the field strength, resulting in a semiconductor-metal transition in the range of 
relatively small electric fields of 200-300 mV \AA$^{-1}$.
These values have been challenged by Liu et al.\cite{QLiu2012}, who reassessed these studies focussing on one material (MoS$_2$), but considered different stacking configurations
of molybdenum and sulphur atoms in the 2D layers. They reported that the electric field strength, at which the band gap closes, is significantly higher, between 1.0 and 1.5 V \AA$^{-1}$, 
and suggested that the smaller values reported by Ramasubramaniam et al.\cite{Ramasubramaniam2011} are caused by applying inappropriate constrains to the symmetry of the bilayer structures.
However, Liu et al. focussed on MoS$_2$ and its band gap without considering SO effects.

\begin{figure}[h!]
\begin{center}
\includegraphics[width=0.45\textwidth,clip]{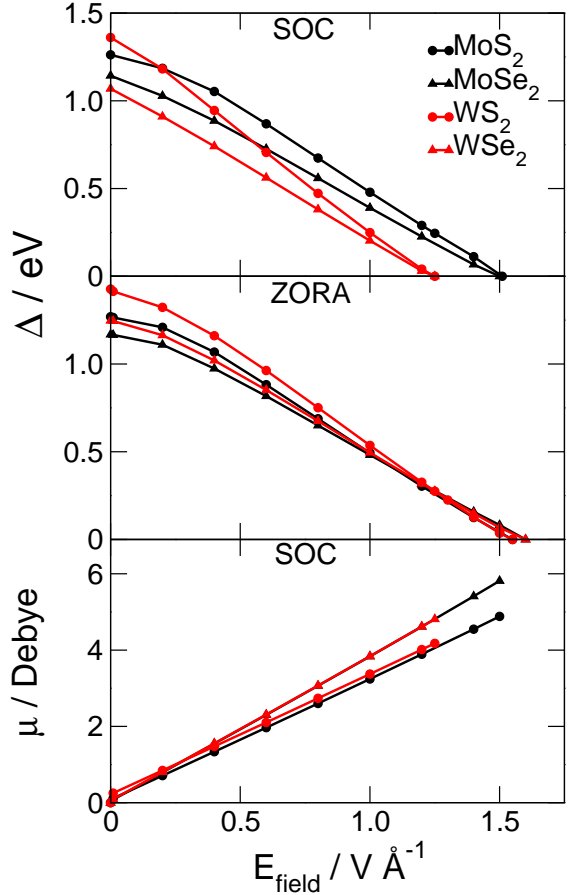}
\caption{\label{fig:bg} Calculated electronic band gaps ($\Delta$) and dipole moments ($\mu$) of TX$_2$ bilayers as function of applied perpendicular electric field. For comparison,
$\Delta$ is reported for both SOC and ZORA calculations.}
\end{center}
\end{figure}

We have calculated the band gap ($\Delta$) evolution with respect to the applied electric field strength for various TX$_2$ bilayers.
Our results, obtained using explicit two-dimensional boundary conditions and thus avoiding possible artifacts due to periodicity in the direction normal to the layers, support the assessment of Liu et 
al.\cite{QLiu2012}: as shown in Fig.~\ref{fig:bg}, the electronic band gaps reduce nearly linearly with applied field strengths and the materials undergo a semiconductor-metal transition at 1.25 V \AA$^{-1}$ 
and 1.50 V \AA$^{-1}$ for WX$_2$ and MoX$_2$ BLs, respectively.
 
In more detail, our calculations agree well with other first-principles simulations where available.
For example, at the equilibrium, MoS$_2$ BL is an indirect band gap semiconductor with $\Delta$ of 1.26 eV in excellent agreement with the calculations of Ramasubramaniam et al.\cite{Ramasubramaniam2011} 
who have obtained 1.26 eV from the fixed experimental structures, and 1.13 eV from structures optimized at the PBE-D2 level. Those values correspond, however, to a different system of higher symmetry, and 
the smaller $\Delta$ values of the optimized structures of Ramasubramaniam et al.\cite{Ramasubramaniam2011} are due to the elongated in-plane lattice vectors (see also Tab. 1).
This is consistent with our earlier work where we have shown that under tensile strain the band gap reduces almost linearly.\cite{Ghorbani2013, Ghorbani2013a}
For the MoS$_2$ BL, Liu et al.\cite{QLiu2012} obtained 1.09 eV band gap at the LDA level and the semiconductor--metal transition at external field of $\sim$1.5 eV \AA$^{-1}$, in close agreement with
our results, but three times larger than those reported by Ramasubramaniam et al.\cite{Ramasubramaniam2011}

Our calculations further show that W-based systems close their band gaps at lower fields than their Mo-based counterparts. This difference is not captured without considering SO effects. 
WS$_2$ bilayers show the strongest band gap dependence on the external field. This system inherently has the largest band gap among all the studied systems, but its $\Delta$ decreases most rapidly with 
the applied field.

\begin{figure}[!h]
\begin{center}
\includegraphics[width=0.75\textwidth,clip]{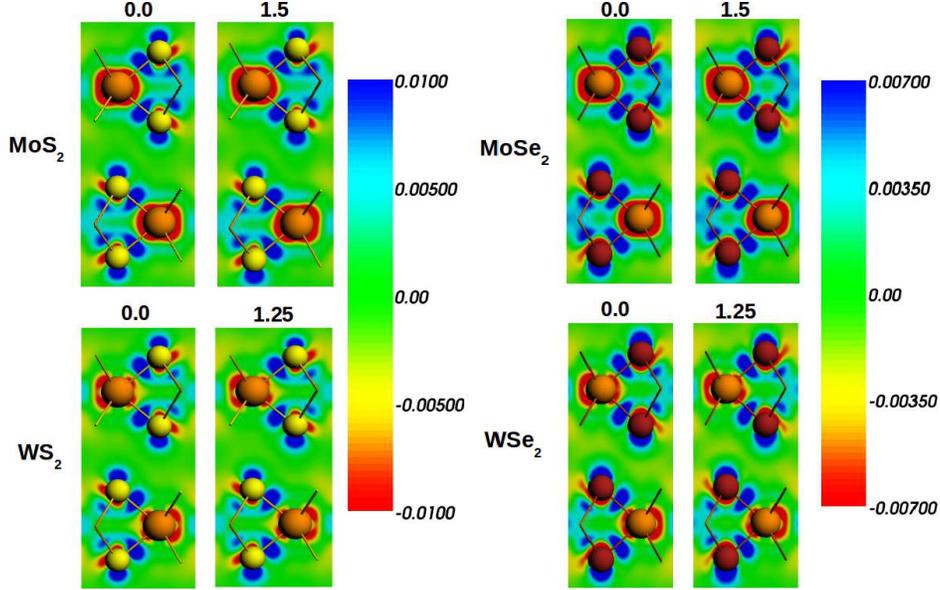}
\caption{\label{fig:dm} Density difference maps calculated at zero and critical electric field strengths for the TS$_2$ (left) and TSe$_2$ (right) bilayers.}
\end{center}
\end{figure}

The polarization of the lattice planes is reflected by the induced dipole moments $\mu$ (see Fig.~\ref{fig:bg}) and the electron density distribution plotted in Fig.~\ref{fig:dm}. 
Dipole moments are significantly larger than for the respective monolayers.\cite{Zibouche2014} Within the field strengths applied in this work, the dipole moments for all systems increase linearly with
the applied field. For critical field strengths, $\mu$ of sulphide BLs are smaller than for selenides by $\sim$1.0 Debye.

\section{Conclusion}

In summary, spin splitting due to the spin-orbit coupling can be induced in centro-symmetric transition metal dichalcogenide bilayers by an external electric field applied perpendicular to the layers. 
The necessary electric fields have a magnitude that can be reached by applying a gate voltage. Thus, the electronic properties of TX$_2$ bilayers can be controlled in a simpler and more effective way 
compared to mechanical deformations.

The electric field causes polarization of individual layers in such a way that the inversion symmetry is broken.
As result, band structures are strongly altered and the spin splitting due to the Stark effect can be enhanced in the valence and conduction bands. The resulting materials are spin- and
valley-polarized semiconductors.  In addition, the electronic band gaps of all TX$_2$ bilayers reduce linearly with applied field and eventually these systems undergo a transition from 
semiconducting to metallic phase at field strengths of 1.2 V \AA$^{-1}$ and 1.5 V \AA$^{-1}$ for W- and Mo-based TX$_2$ bilayers, respectively.
As such field strengths could be realized in practical nanoelectronic devices, we expect very interesting application possibilities in the emerging field of spin- and valleytronics.

\section{Acknowledgements}
Financial support by Deutsche Forschungsgemeinschaft (DFG, HE 3543/17-1) and the European Commission through the Initial Training Network (ITN) MoWSeS (GA FP7-PEOPLE-2012-ITN)
and the Industrial Academic Partnership Pathways (IAPP) QUASINANO (GA FP7-PEOPLE-2009-IAPP) is acknowledged.

\end{document}